# Band gap opening at the Dirac point in Co/BiSbTeSe$_2$(0001) system

A. K. Kaveev[1], A.G. Banshchikov[1], A. N. Terpitskiy[1], V. A. Golyashov[2], O. E. Tereshchenko[2,3], K.A. Kokh[4], D.A. Estyunin[5], A.M. Shikin[5] and E.F. Schwier[6]

[1]*Ioffe Institute, 194021 Saint - Petersburg, Russia*
[2] *Rzhanov Institute of Semiconductor Physics, 630090 Novosibirsk, Russia*
[3]*Novosibirsk State University, 630090 Novosibirsk, Russia*
[4]*Sobolev Institute of geology and mineralogy, 630090 Novosibirsk, Russia*
[5]*Saint-Petersburg State University, 198504, Saint-Petersburg, Russia*
[6]*Hiroshima University, 739-8527, Higashi Hiroshima, Japan*

**Abstract**. Sub-angstrom Co coverage, being deposited on BiSbTeSe$_2$(0001) surface at 200-330°C, opens a band gap at the Dirac point, with the shift of the Dirac point position caused by RT adsorbate pre-deposition. Temperature dependent measurements in 15-150 K range have shown no band gap width change. This fact indicates the nonmagnetic nature of the gap which may be attributed to the chemical hybridization of surface states upon the introduction of Co adatoms, which decrease crystallographic symmetry and eliminate topological protection of the surface states.

## 1. Introduction

To date a novel class of materials – topological insulators attracts essential interest in the area of spintronics, because of their exotic properties. These materials demonstrate insulating properties in the bulk. At the same time their surface possesses conduction, owing to a strong spin–orbit interaction. This interaction results in the spin splitting of the surface states [1]. These states are linearly dispersed and protected by time reversal or crystalline symmetry, forming a Dirac cone near the Dirac point [1, 2]. So-called "spin-momentum locking" related to chiral spin structures represents an important feature of these states [3]. The current flow through these states is resistant to the scattering on the surface defects like non-magnetic impurities, scratches etc, being "topologically protected". Studies of the ferromagnetic metal (FM) – topological insulator (TI) interface is attractive from the point of view of the spin-to-charge conversion, the mutual control of the magnetization and current through topological states, which is prospective for the development of spintronic devices based on the control of the current through the topological states [4, 5]. It is known [6] that one may observe the magnetic proximity effect at this interface, which accounts for breaking of time reversal or crystalline symmetry in topological insulators. This effect shows in the exchange field influence on the spin of the topological states, which conducts to the removal of spin degeneracy in the Dirac point and the energy gap opening. And vice versa, the spin-polarized current flow through the topological states of Tis may influence on the magnetic order in FMs, like spin transfer and spin orbit torque [7, 8]. From the point of view of the mentioned application of epitaxial FM/TI systems, it is important to study modification of the TI's electronic, structural and interface properties under FM material deposition. There are a few works related to experimental studies of the contact between FM and TI in different FM/TI systems, for example, with the use of Bi$_2$Se$_3$ [9] or Sb$_2$Te$_3$ [10, 11] as TI. In the present work, the studies of transformation of the BiSbTeSe$_2$ (hereinafter – BSTS2) structure and electronic states during Co deposition were carried out.

## 2. Experimental

Co was deposited on BSTS2(0001) surface with use of molecular beam epitaxy (MBE). Crystalline topological insulator BSTS2 substrates were obtained by modified Bridgeman method. Clean TI surface was obtained using cleavage with scotch tape and subsequent vacuum annealing at 330°C during 30 minutes under $10^{-8}$ mBar pressure. Co adatoms were deposited on

BSTS2 at the temperature range from the room temperature (RT) to 330°C in vacuum. The effective thickness of the material was in 0 - 2 Å range. The electronic structure of the grown samples was studied by the laser angle-resolved photoelectron spectroscopy (Laser ARPES) method [12]. The measuring system is based on a 3.3 W Ti:Sa mode-locked laser (Tsunami, Spectra Physics) with a 17 W green laser pumping (Millenia eV, Spectra Physics). High harmonic generator (HarmoniXX, A. P. E.) allows one to obtain UV radiation at a 191-210 nm wavelength with a power of 0.1-0.7 mW. To measure the angular dependence of the energy of photoelectrons, a spectrometer with an angular resolution in two coordinates was used. The Laser ARPES method is a powerful tool for studying the dispersion law and features of the surface band structure. The use of high-energy laser radiation with external focusing allows us to achieve a uniquely high (~ 5 micrometer) spatial resolution over the sample surface compared to the synchrotron radiation commonly used in ARPES systems (around 100-1000 micrometers). To search the Dirac point position, reciprocal space ($k_x$, $k_y$) mapping around the Dirac point was carried out.

## 3. Results and discussion

The BSTS2 (0001) surface was prepared by cleavage in ultrahigh vacuum (~ $10^{-9}$ mbar). Using the Laser ARPES method, it was shown that a clean BSTS2 surface demonstrates the presence of surface states with a linear Dirac-type dispersion law, while the Fermi level lies in the bulk band gap and intersects surface states at the Dirac point. With such surface state structure, the observation of its changes near the Dirac point by the Laser ARPES method is impossible. However, it was found that deposition of the ultra-small amounts of adsorbate (about 0.1 Å) at room temperature leads to band bending with a shift of the Dirac point below the Fermi level (Fig.1).

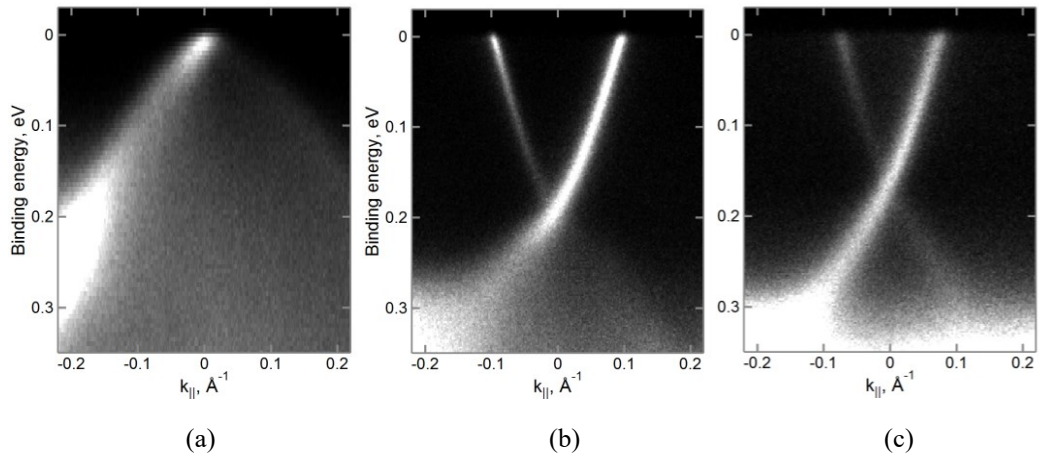

(a)          (b)          (c)

Fig. 1. (a) - electronic structure of BSTS2 surface, measured using Laser ARPES (hv = 6.3 eV, T = 20 K), after cleaving the sample in vacuum. (b, c) - bending of the surface bands after deposition of 0.1 Å of Co (b) or Mn (c) at room temperature.

In our case the role of adsorbate is played by Co or Mn. Note that no Dirac point shift was observed in case of the temperature of adsorbate deposition increase (Fig. 2 (a)).

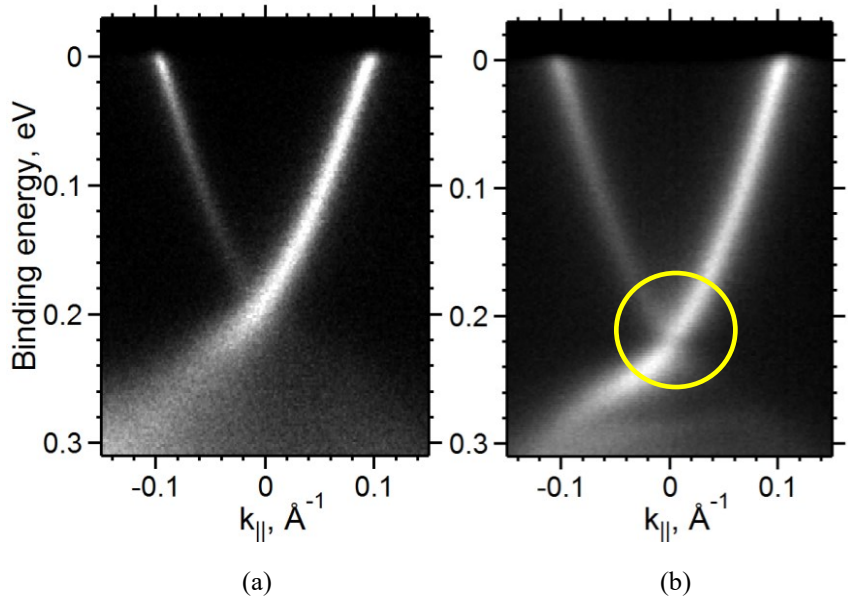

(a) (b)

Fig. 2. The band bending after deposition of 0.1 Å adsorbate at room temperature (a) and the band gap opening (b) in the spectrum of Dirac surface states in the Co / BSTS2 system. Figure (b) corresponds to the deposition of 0.6 Å of Co at 300 °C.

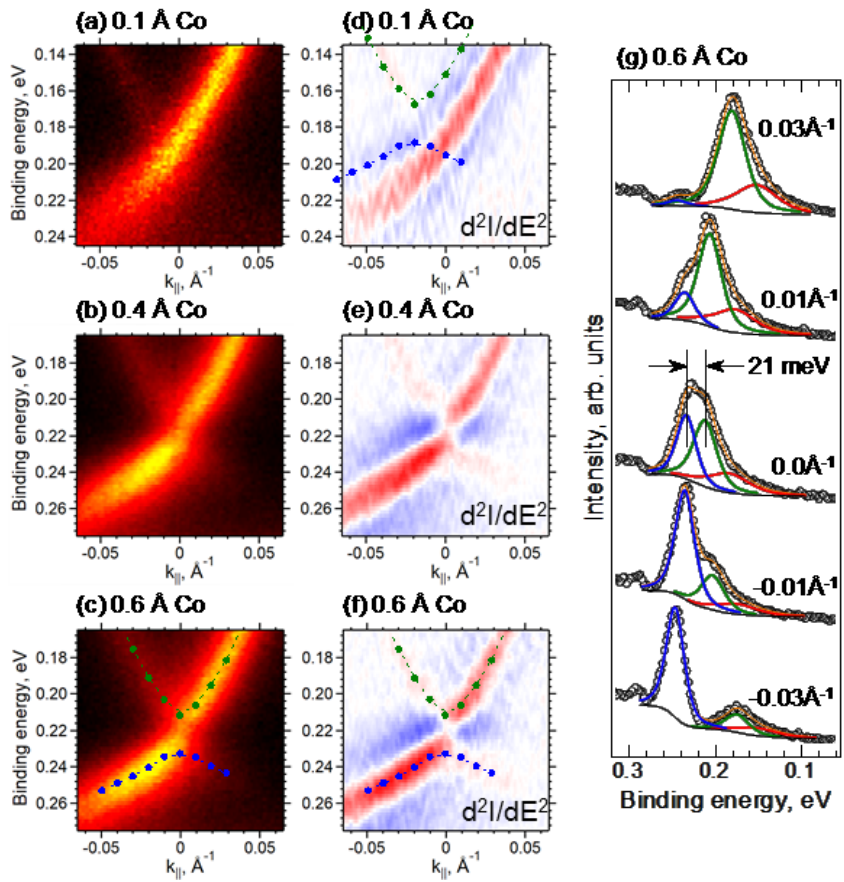

Fig. 3. ARPES spectra of BSTS2 topological surface states near Dirac point for samples with 0.1 Å (a), 0.4 Å (b) and 0.6 Å (c) of Co deposited with their $d^2I/dE^2$ second derivatives (d-f) respectively. (g) EDC intensity profiles for different $k_\parallel$ values of spectrum (c) in this figure (0.6 Å of Co). The data were fitted using two Voight profiles for the lower (blue curve) and the upper (red curve) Dirac cone states and a Shirley background, third Voight profile (green curve) was used to account for asymmetry of the upper Dirac cone state. Derived energy positions of the fitted peaks are plotted as points and dashed lines in (c) and (f).

Further cobalt deposition results in the band gap opening in the Dirac surface state spectrum (Fig. 2 (b)). To show the gap in more details, we have calculated the second derivatives $d^2I/dE^2$ of the image and fitted the energy distribution curves (EDC) cuts at various $k_\parallel$ values using Voight profiles. Fig. 3 shows Laser ARPES spectra of BSTS2 topological surface states near Dirac point for samples with 0.1 Å (a), 0.4 Å (b) and 0.6 Å (c) of Co deposited with their $d^2I/dE^2$ second derivatives (d-f) respectively. The energy gap near the Dirac point is clearly seen. Energy distribution curve intensity profiles cut at different $k_\parallel$ values for 0.6 Å of Co are shown in Fig. 3 (g). The data in (g) were fitted using two Voight profiles for the lower and the upper Dirac cone states and a Shirley background. This background may be attributed to Co-contained crystalline phase. The third Voight profile was used to account for asymmetry of the upper Dirac cone state. Derived energy positions of the fitted peaks are also shown in the Fig. 3 (c, f). These fit results allow estimation of the gap band width as 21 meV.

The existence of the band gap was observed for Co coatings from 0.4 Å and more, at least up to 2 Å (see Fig. 4, the spectra for different Co amount), which is approximately equal to the cobalt lattice constant (2.5 Å).

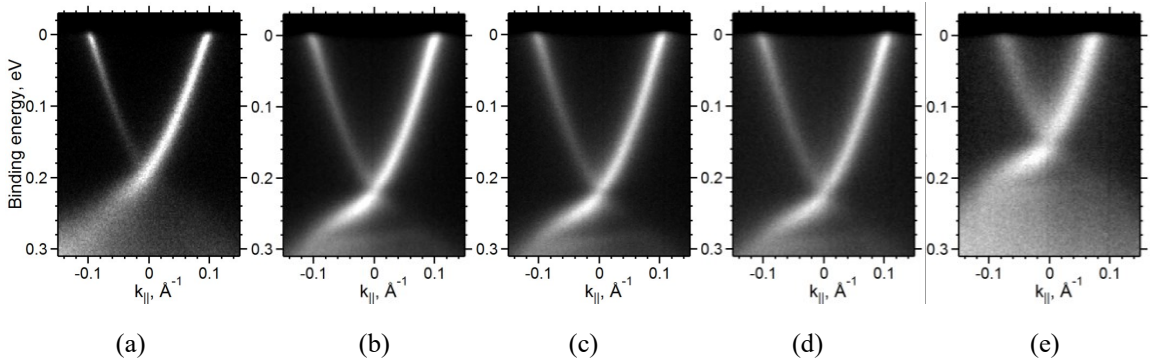

(a) (b) (c) (d) (e)

Fig. 4. (a-d) - changes in electronic structure of BSTS2 surface during step-by-step deposition of total 0.1 Å (a), 0.4 Å (b), 0.6 Å (c) and 1 Å (d) of Co at substrate temperature of 300 ºC and total 2 Å (e) of Co with subsequent anneal at 325 ºC. The energy gap, appeared at Dirac point, can be clearly seen in all the spectra.

The band gap width was found to be constant, not depending on the cobalt amount. Moreover, it was shown that the gap opening by Co deposition apparently doesn't depend on the adsorbate composition. Fig. 5 (a-d) shows the Laser ARPES spectra for 0.4 Å of Co deposited on the BSTS2 substrate at room temperature (c) with preliminary deposited 0.1 Å of Mn adsorbate (b) to provide the Dirac point shift relative to its position for clear surface (a). The gap opening is presented as in the case of pre-deposition of the Co adsorbate (see Fig. 2)), so for the Mn adsorbate (Fig. 5 (c, d)).

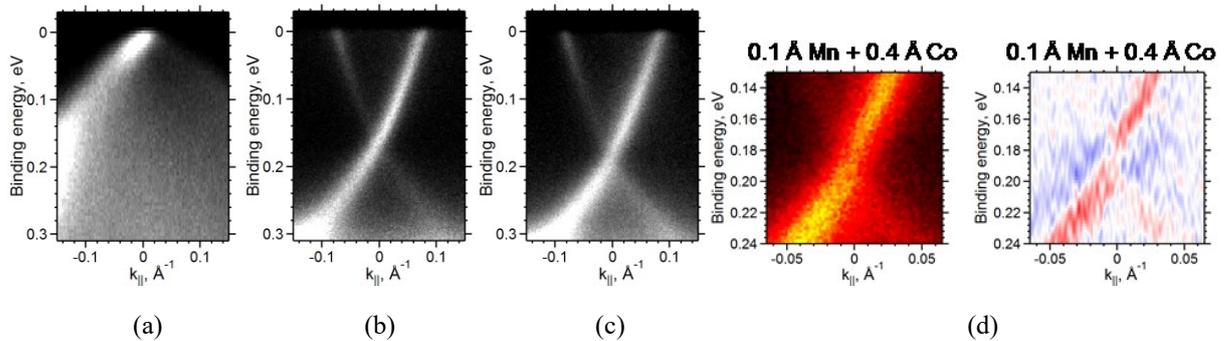

(a) (b) (c) (d)

Fig. 5. (a) - electronic structure of BSTS2 surface, measured using Laser ARPES (hv = 6.4 eV, T = 20 K), after cleaving the sample in vacuum, (b) deposition of 0.1 Å of Mn at room temperature and (c) consequent deposition of 0.4 Å of Co at room temperature. (d) – zoom of (c) and its $d^2I/dE^2$ second derivative.

To clarify the nature of the band gap, preliminary measurements of the dependence of its width on temperature in the 15-150 K range were carried out. No change in the width of the gap was noticed, which suggests the nonmagnetic nature of the gap. Magnetic and nonmagnetic gaps were found, respectively, in [13, 14] when studying the bulk TI $Bi_2Se_3$ doped with Mn in the first work, and during MBE growth with intercalation of Mn ions in $Bi_2Se_3$ in the second work. There are theoretical works which explain the presence of a nonmagnetic gap with chemical hybridization of surface states upon the introduction of impurity atoms [15], which can also occur in case of Co deposition. Hybridization or another distortion eliminates topological protection of surface states because of the top layer crystallographic symmetry decrease [16]. Besides the magnetic field application, the distortion represents another way to tune the surface band gap, which is useful for electronic device applications. We assume dissolution of Co in the top part of BSTS2 substrate as a way to the hybridization and distortion.

In summary, it was found that being deposited on BSTS2(0001) surface with 0.4-2 Å coating effective thickness, Co opens band gap near the Dirac point. The approximate gap width is about 21 meV. According to the temperature dependent LARPES measurements the gap has nonmagnetic nature. It should be noted that there is an extremely small number of works devoted to the experimental observation of Me / TI systems with opened band gap near the Dirac point. The opening of the gap observed in these works (as a rule, using Mn atoms) occurs in case of TI doping by ferromagnetic metal. At the same time, there are no works associated with the opening of the energy gap in the Me / TI system during the deposition of metal on the TI surface. In particular, previous attempts to open the energy gap by depositing Fe and Co on the surface of the $Bi_2Se_3$ and $Bi_2Te_3$ were unsuccessful [19-21].

*Acknowledgement*

This work has been supported by Russian Foundation for Basic Research (grant no 17-02-00729). The ARPES measurements at HiSOR were performed with the approval of Proposal Assessing Committee (Proposal Numbers: 19AG016).